\documentclass[twocolumn,aps,prl,groupedaddress,amsmath,amssymb,longbibliography]{revtex4-2}
\usepackage{url}
\usepackage{ulem}
\usepackage[colorlinks=true, linkcolor=blue,urlcolor=blue,anchorcolor=blue,citecolor=blue,bookmarksnumbered]{hyperref}
\usepackage{graphicx}
\usepackage{dcolumn}
\usepackage{bm}
\usepackage{verbatim}

\begin{document}
	
	\title{Accelerated Topological Pumping in Photonic Waveguides\\ Based on Global Adiabatic Criteria}
	\author{Kai-Heng~Xiao$^{1}$}\email[]{These authors contributed equally to this work.} \author{Shi-Lei~Su$^{2}$}\email[]{These authors contributed equally to this work.}\author{Xiang~Ni$^{3}$}\email[]{These authors contributed equally to this work.} \author{Yi-Ke~Sun$^{1}$}\author{Jin-Kang~Guo$^{3}$} \author{Zhi-Yong~Hu$^{1}$} \author{Xu-Lin~Zhang$^{1}$} \author{Jia~Li$^{2}$} \author{Jin-Lei~Wu$^{2}$}\email[]{jlwu517@zzu.edu.cn} \author{Zhen-Nan~Tian$^{1}$}\email[]{zhennan\_tian@jlu.edu.cn} \author{Qi-Dai~Chen$^{1}$}\email[]{chenqd@jlu.edu.cn}
	
	\affiliation{$^{1}$State Key Laboratory of Integrated Optoelectronics, JLU Region, College of Electronic Science and Engineering, Jilin University, Changchun 130012, China\\
		$^2$School of Physics, Zhengzhou University, Zhengzhou 450001, China\\
		$^3$School of Physics, Central South University, Changsha 410083, China}
	
	\begin{abstract}		
		Adiabatic topological pumping enables robust transport of energy and information, yet its operational speed is fundamentally constrained by the instantaneous adiabatic condition, which necessitates prohibitively slow parameter variations. Here, we propose a paradigm shift from instantaneous to global adiabaticity. We derive a global adiabatic criterion~(GAC) that establishes an absolute fidelity bound by controlling the root-mean-square nonadiabaticity. Building on this framework, we introduce a fluctuation-suppression acceleration criterion to minimize spatial inhomogeneity, allowing for a safe increase in mean nonadiabaticity without compromising fidelity. We experimentally demonstrate this principle in femtosecond-laser-written photonic Su-Schrieffer-Heeger waveguide arrays via scalable power-law coupling modulation. Our accelerated topological pumping achieves a fidelity of $>0.95$ with a fivefold reduction in device length compared to conventional schemes, exhibits the predicted linear scaling with system size, and maintains robust performance across a bandwidth exceeding $400$~nm. This GAC framework provides a universal design rule for fast, compact, and robust adiabatic devices across both quantum and classical topological platforms.
	\end{abstract}
	
	\maketitle
	\textbf{\textit{Introduction}}---Topological photonics has emerged as a groundbreaking paradigm, offering robust energy transport immune to structural imperfections and disorder~\cite{Lu2014NatPhoton,Noh2018NatPhoton,Ozawa2019RMP,Ni2023CR}. This robustness arises from the topological protection of photonic edge states, which enables high-efficiency energy propagation even in the presence of defects, scattering, or fabrication variations~\cite{Szameit2024NP,Wang2024PI,Gao2024npjN,Yan2024npjN}. Notably, the well-known Su-Schrieffer-Heeger~(SSH) model~\cite{SSH1979PRL,SSH1980PRB} has become a cornerstone for studying phenomena such as topological pumping, in which directed energy transfer is achieved through adiabatic modulation of system parameters~\cite{Kraus2012,Lohse2018Nature,Zilberberg2018Nature,FMei2018PRA,Jurgensen2021Nature,Sun2022NP,Cheng2022NC,Citro2023NRP,Walter2023NP}. Recent experiments have demonstrated that SSH-based topological pumping across various platforms can facilitate energy transport using diverse physical carriers~\cite{TTian2022PRL,Deng2022Science,CHWu2023PRA,TTian2024PRB,ZGuan2024PRApplied,Song2024SciAdv,WJLiu2022PRA,Wu2024NC}.

	However, the practical implementation of such systems is fundamentally constrained by adiabatic protocols, which demand prohibitively large spatial footprints and conflict with the ongoing push toward device miniaturization~\cite{Garanovich2012PR,Enrich2016RPP,Vitanov2017RMP,Privitera2018PRL}. While various strategies have been explored to accelerate topological pumping, including shortcut engineering techniques~\cite{Odelin2019RMP,Angelis2020PRA,Romero2024PRApplied,Guo2024PRA,JXu2024PRA,WLiu2025PRA}, structure reforming~\cite{Longhi2019PRB,TTian2022PRL,Song2024SciAdv,Deng2022Science,CHWu2023PRA,TTian2024PRB,ZGuan2024PRApplied,Zurita2023Quantum}, and coupling tailoring~\cite{Brouzos2020PRB,XDZhao2023NPJQI,JXHan2024PRApplied,Palaiodimopoulos2021PRA,WJLiu2022PRA,Wu2024NC}, they often optimize only local segments of the evolution, exacerbate experimental complexities, or increase susceptibility to parameter fluctuations. These limitations stem from the commonly adopted instantaneous adiabatic condition, which imposes a pointwise constraint while overlooking the global statistical properties of evolution. Consequently, existing approaches fail to provide a holistic design framework capable of simultaneously ensuring fast, high-fidelity transport across the entire pumping trajectory while maintaining scalability and robustness. 
	
	In this work, we introduce a paradigm shift from instantaneous to \textit{global} adiabaticity. We theoretically show that high fidelity is governed not by suppressing nonadiabaticity at every point, but by controlling its root-mean-square (RMS) value over the entire evolution. Based on this insight, we develop two global accelerated adiabatic criteria~(GAAC): a global adiabatic criterion~(GAC) that enforces an absolute fidelity bound, and a fluctuation-suppression acceleration criterion~(FSAC) that minimizes penalties arising from spatial inhomogeneity. This framework fundamentally reformulates the conventional adiabatic constraint, permitting local rapid parameter variations in topologically protected region, thereby enabling accelerated, linearly scalable pumping. We experimentally validate this paradigm in a femtosecond-laser-written photonic SSH lattice. By designing coupling profiles that satisfy the GAAC, we achieve high-fidelity topological pumping with a fivefold reduction in device length compared to conventional adiabatic scheme. In addition, the accelerated topological pumping exhibits exceptional robustness over a bandwidth exceeding $400$~nm. Beyond topological pumping, the  GAAC principle provide a general strategy for accelerating any adiabatically driven quantum and classical evolutions, including quantum gate operations~\cite{JWu2025NPJQI,ZZhu2025PRL}, Landau-Zener transitions~\cite{ZGChen2021PRL,Bjorkman2025PRL}, and higher-dimensional topological pumping protocols~\cite{Sun2024NC,PYSong}.

	\textbf{\textit{Theory framework}}---We start from the Schr\"{o}dinger-like coupled-mode equation~(CME) defined with respect to the spatial coordinate $z$ (consistent with our photonic waveguide platform) $\hat{H}(z)|\Psi(z)\rangle = i \partial_z |\Psi(z)\rangle$, where $|\Psi(z)\rangle$ is the system state vector. For a topological pumping system, we assume that the Hamiltonian $\hat{H}(z)$ possesses a non‑degenerate topological gap mode $|\mathrm{GAP}(z)\rangle$ with eigenenergy $E_0(z)$, separated from the $m$th bulk mode $|\phi_m(z)\rangle$ ($m\neq 0$) by a finite energy gap $\Delta E_m(z) = |E_m(z)-E_0(z)|>0$. The conventional approach relies on the instantaneous adiabatic condition: $Q_m(z)\equiv\left| \frac{\langle \phi_m(z) | \partial_z | \mathrm{GAP}(z) \rangle}{\Delta E_{m}(z)}\right|\ll 1$, $\forall z\in[0,L]$~\cite{Vitanov2017RMP,Enrich2016RPP}. This pointwise constraint enforces uniformly slow parameter variations, leading to prohibitively long device lengths.
	
	Here we define the instantaneous nonadiabaticity factor as the sum over all bulk modes $Q_{\mathrm{non}}(z) \equiv \sum_{m\neq 0} Q_m(z)$. Then we introduce a paradigm‑shifting framework that replaces the instantaneous adiabatic condition with GAAC:
	
	(i)~Global adiabatic criterion [see Note~1 in Supplemental Material~(SM)~\cite{SM} for detailed derivation]:
	\begin{equation}
	L \Delta E_{\text{min}}\sqrt{ \overline{Q}_{\rm non}^2 + \sigma_Q^2 } \leq \sqrt{\varepsilon_c}.
	\label{Criterion1}
	\end{equation}
	
	(ii)~Fluctuation-suppression acceleration criterion:
	\begin{equation}
	\sigma_Q \ll {\overline Q}_{\mathrm{non}}.
	\label{Criterion2}
	\end{equation}\\
	$L$ is the evolution length. ${{\overline Q}_{\mathrm{non}}} = (1/L)\int_0^L Q_{\mathrm{non}}(z) \mathrm{d}z$ is defined as the spatial mean of the nonadiabaticity factor. $\sigma_Q^2 = 1/L\int_0^L \delta Q(z)^2 \mathrm{d}z$ is the variance ($\sigma_Q$ the standard deviation), with the fluctuation $\delta Q(z) = Q_{\mathrm{non}}(z) - {\overline Q}_{\mathrm{non}}$. $\Delta E_{\text{min}}=\min_{z,m}[\Delta E_m(z)]$ is the minimum energy gap along the whole evolution. $\varepsilon_c= 1 - {\mathcal F}_c$ is the maximum allowed infidelity with ${\mathcal F}_c$ being the critical value of the target fidelity.  
	
	GAC in Eq.~\eqref{Criterion1} sets an absolute upper bound on the product of device length and the RMS value of nonadiabaticity factor, thereby guaranteeing the desired fidelity. FSAC in Eq.~\eqref{Criterion2} eliminates the ``penalty term'' arising from fluctuation, in the cumulative nonadiabatic coupling by minimizing spatial inhomogeneity, thereby permitting a larger mean driving strength ${\overline Q}_{\mathrm{non}}$ without violating the adiabatic bound. When FSAC is satisfied, GAC simplifies to  $L \, {\overline Q}_{\mathrm{non}} \, \Delta E_{\mathrm{min}} \leq \sqrt{\varepsilon_c}$.
	It reveals the central acceleration mechanism: for a given target fidelity and system gap, increasing the mean nonadiabaticity factor ${\overline Q}_{\mathrm{non}}$ (i.e., permitting faster parameter variations) directly shortens the required device length. Moreover, because the minimum gap $\Delta E_{\mathrm{min}}$ typically decreases with increasing system size $N$, our criteria inherently govern the scaling of $L$ with $N$.
	
\begin{figure}[t]\centering
		\includegraphics[width=\linewidth]{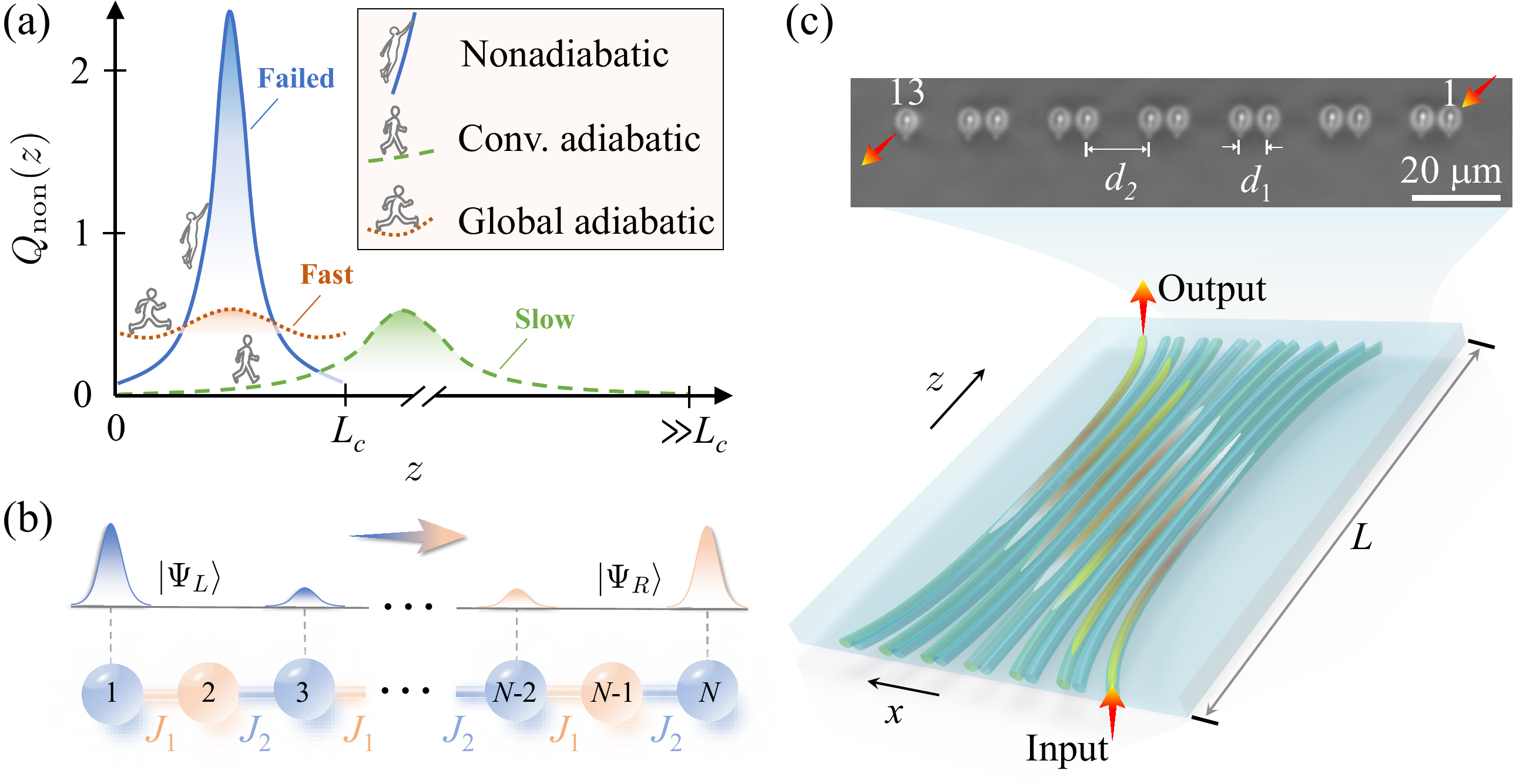}
		\caption{\textbf{Illustration of acceleration strategy.} (a)~Schematic diagram of the nonadiabatic factor $Q_{\mathrm{non}}(z)$ affecting topological pumping. The global adiabatic scheme with a higher and flatter $Q_{\mathrm{non}}(z)$ enables fast topological pumping, while the conventional scheme (labeled ``conv.") is slow with lower ${\overline Q}_{\mathrm{non}}$ or failed with higher ${\overline Q}_{\mathrm{non}}$. (b)~Dimerized SSH model with modulated intracell and intercell hopping rates $J_1(z)$ and $J_2(z)$, implementing topological transport from the left edge state $|\Psi_L\rangle$ to the right $|\Psi_R\rangle$. (c)~Bottom panel: schematic of a spatially modulated SSH lattice of photonic waveguides fabricated inside a glass; Top panel: photograph of the fabricated sample of waveguides at the output facet.}\label{f1}
	\end{figure}

	To illustrate our principle, Fig.~\ref{f1}(a) schematically depicts the proposed acceleration strategy for topological pumping. The nonadiabaticity $Q_{\mathrm{non}}(z)$ satisfying GAAC evolves more uniformly, maintaining a relatively large mean value while exhibiting reduced fluctuations. This profile enables fast topological pumping to achieve target fidelity ${\mathcal F}_c$ within a critical evolution length $L_c$. In contrast, conventional adiabatic designs require a substantially longer device length to reach the same fidelity, and fail to do so when compressed to $L_c$ due to enhanced nonadiabatic errors.
	
	\textbf{\textit{System and acceleration strategy}}---We execute exemplarily our investigation with the renowned dimerized SSH model [bottom panel in Fig.~\ref{f1}(b)]. In experiment, we fabricate the photonic waveguides~(odd-numbered quantity is $N$) to construct the SSH model within a boroaluminosilicate glass by femtosecond-laser direct writing techniques~[Fig.~\ref{f1}(c)]~\cite{Zhang2022NatPhoton,Sun2022NP,ACrespi2023NanoP,JLWu2025NC}. This waveguide array with the length $L$~(maximum $75$~mm for our chips) is constrained to nearest-neighbor couplings, with staggered strengths $J_1$~(intracell) and $J_2$~(intercell) being spatially modulated in the $z$-direction.
	
	Under the tight-binding approximation, we define the state vector $|\Psi(z)\rangle=\sum_n^N{\mathcal C}_n(z)|n\rangle$ of light in the waveguides, where ${\mathcal C}_n(z)$ is the complex probability amplitude of basis vector $|n\rangle$ on the $n$th waveguide, satisfying normalization $\sum_n^N|{\mathcal C}_n(z)|^2=1$. The SSH Hamiltonian reads
	\begin{equation}{\hat H}(z)=\sum_{j=1}^{(N-1)/2}J_1(z)|2j-1\rangle\langle2j|+J_2(z)|2j+1\rangle\langle2j|+{\rm H.c.}
	\end{equation}
	When the waveguide array is adiabatically modulated along the $z$ direction~\cite{Enrich2016RPP,Vitanov2017RMP,FMei2018PRA,Jurgensen2021Nature,Song2024SciAdv,WJLiu2022PRA,Wu2024NC}, the input light can be topologically pumped from one edge waveguide to the other following the spatial profiles of the zero-energy gap modes $|{\rm GAP}(z)\rangle= {\mathcal N}\sum_{j=1}^{(N-1)/2}\beta^{j}(z)|2j+1\rangle$~(Note~2 in SM~\cite{SM}), where $\beta(z)\equiv-J_1(z)/J_2(z)$ and ${\mathcal N}$ denotes the normalization. For the topologically nontrivial~(trivial) phase under $J_1<J_2$~($J_1>J_2$) (Note~3 in SM~\cite{SM}), the gap mode is localized on the left~(right), as shown by $|\Psi_{L(R)}\rangle$ in Fig.~\ref{f1}(b) (top panel). The topological pumping will be triggered when the waveguide array is adiabatically modulated from $J_1(z=0)\ll J_2(z=0)$ to $J_1(z=L)\gg J_2(z=L)$.
	
	\begin{figure}\centering
		\includegraphics[width=\linewidth]{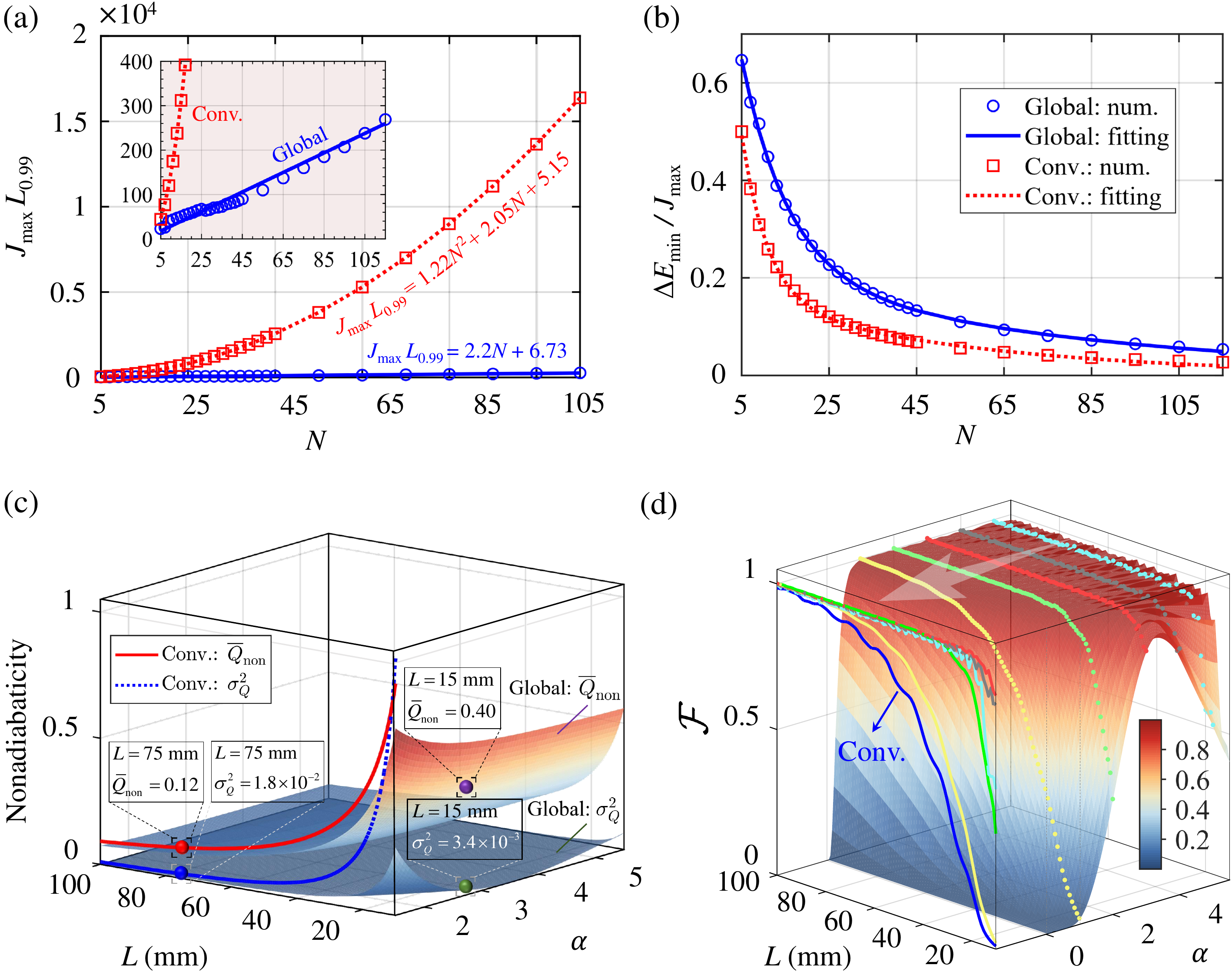}
		\caption{\textbf{Scalability of GAAC-based topological pumping}. (a)~Critical device length $L_{0.99}$ (in units of $J_{\text{max}}^{-1}$) required to achieve $\mathcal{F}(L) >0.99$, $\forall L>L_{0.99}$, as a function of $N$. For our strategy (blue), circles: numerical data; Solid line: fitted data. For the conventional scheme (red), squares: numerical data; Dashed line: fitted data. (b)~Comparison of minimum energy gaps between our acceleration strategy and the conventional scheme~(see Note~4 in SM~\cite{SM} for more details). (c)~${\overline Q}_{\mathrm{non}}$ and $\sigma_Q^2$ versus $L$ and $\alpha$ for $N=13$ based on the global adiabatic criteria and the conventional adiabatic scheme, respectively. (d)~Numerical fidelity versus different $L$ and $\alpha$ for $N=13$, calculated by CME. Five ${\mathcal F}$-$L$ lines with $\alpha=1$-$5$, as well as the ${\mathcal F}$-$L$ line of the conventional pumping, are projected onto the leftmost ${\mathcal F}$-$L$ plane for comparison.}\label{f2}
	\end{figure}

	To match our GAAC, here we take a power-law envelopes of inter-waveguide couplings
	$J_1(z)=J_{\max}\left[1-\left(\frac{L-z}{L}\right)^{\alpha(N)}\right],\quad$ and $J_2(z)=J_{\max}\left[1-\left(\frac{z}{L}\right)^{\alpha(N)}\right]$ as a representative example, which provide the global adiabatic $Q_{\mathrm{non}}(z)$ in Fig.~\ref{f1}(a). The $N$-dependent dimensionless parameter $\alpha(N)=1.66\ln N-1.2$ is determined by numerical fitting methods~(Note~4 in SM~\cite{SM}). In Fig.~\ref{f2}(a) we show the scalability of our acceleration strategy by calculating the critical lengths $L_{0.99}$ that guarantees pumping fidelity ${\mathcal F}(L)>0.99$, $\forall L>L_{0.99}$, for which the fidelity is ${\mathcal F}(z)=|{\mathcal C}_N(z)|^2$ with initialization ${\mathcal C}_1(z=0)=1$. The increase of $L_{0.99}$ with $N$ follows a linear trend fitted by~(Note~4 in SM~\cite{SM})
	\begin{equation}
	J_{\rm max}L_{0.99}=2.2N+6.73.
	\end{equation}
	In contrast, it is a quadratic function $J_{\rm max}L_{0.99}=1.22N^2+2.05N+5.15$ for the conventional pumping scheme based on $J_1(z)=0.5J_{\max}\left[1-\cos({\pi z}/{L})\right]$ and $J_2(z)=0.5J_{\max}\left[1+\cos({\pi z}/{L})\right]$~\cite{FMei2018PRA,Jurgensen2021Nature,JXHan2024PRApplied,Song2024SciAdv,Wu2024NC,WJLiu2022PRA}. This pair of coupling strengths yields the conventional adiabatic and nonadiabatic profiles of $Q_{\mathrm{non}}(z)$ shown in Fig.~\ref{f1}(a). This sharp contrast concerning the $L_{0.99}$-$N$ relationship underlines significant acceleration and scalibility superiority of our strategy over the conventional one. Distinct enlargement of $\Delta E_{\rm min}$ [Fig.~\ref{f2}(b)] in our acceleration strategy is an important factor resulting in a much shorter $L_{0.99}$ than the conventional scheme, which can further decrease the upper bound of evolution length according to GAC in Eq.~\eqref{Criterion1}.
	
	To connect the $N$-dependent adaptive parameter $\alpha(N)$ with our GAAC, we take $N=13$ as an example~(See Note~5 in SM~\cite{SM} for sample fabrication). Figure~\ref{f2}(c) compares ${\overline Q}_{\mathrm{non}}$ and $\sigma_Q^2$ between the conventional scheme and our strategy. On the one hand, our strategy achieves significant suppression of both ${\overline Q}_{\mathrm{non}}$ and $\sigma_Q^2$ compared to the conventional approach, which is attributed to overall reduction of global nonadiabaticity. On the other hand, while ${\overline Q}_{\mathrm{non}}$ decreases with increasing $\alpha$ (with a diminishing rate of decrease), $\sigma_Q^2$ exhibits a clear minimum around $\alpha=3$. Referring to Fig.~\ref{f2}(d), the peak in pumping fidelity identifies an optimal value around $\alpha=3$, which enables a significantly shorter device length to achieve the same target fidelity. Under the GAAC design with $\alpha(13)=3$, the critical length required to achieve ${\mathcal F}(L)>0.95$ is $L_{0.95}=15$~mm, $\forall L>L_{0.95}$. In contrast, the conventional scheme requires a length of 75~mm to reach the same fidelity, indicating a {\it fivefold} reduction via our scheme in device footprint.

	Existing schemes~\cite{Brouzos2020PRB,XDZhao2023NPJQI,Brouzos2020PRB,JXHan2024PRApplied,Palaiodimopoulos2021PRA,WJLiu2022PRA,Wu2024NC} for accelerating topological pumping may partially align with our GAAC. However, these schemes either do not fully satisfy both criteria simultaneously or lack a comprehensive framework that guarantees performance (speed and robustness) and scalability~(via $N$-dependent parameters).
	Specially, we focus on comparison with recently reported fast topological pumping schemes~(Note~6 in SM~\cite{SM}), such as the coherent adiabatic passage with $J_{\rm max}L|_{N=31}=400$ for ${\mathcal F}>0.99$~\cite{Longhi2019PRB,TTian2022PRL}, the quantum metric with $J_{\rm max}L|_{N=8}=44.6$ for ${\mathcal F}>0.9$~\cite{Song2024SciAdv}, and the adiabatic infimum with $J_{\rm max}L|_{N=8}=56.8$ for ${\mathcal F}>0.9$~\cite{Wu2024NC}. Our acceleration strategy for implementing ${\mathcal F}>0.99$ requires even more compact footprints $J_{\rm max}L|_{N=31}=70$ and $J_{\rm max}L|_{N=9}=40$. This remarkable contrast clearly demonstrates the superiority of our strategy in terms of topological pumping speed.
	
	\begin{figure}\centering
		\includegraphics[width=\linewidth]{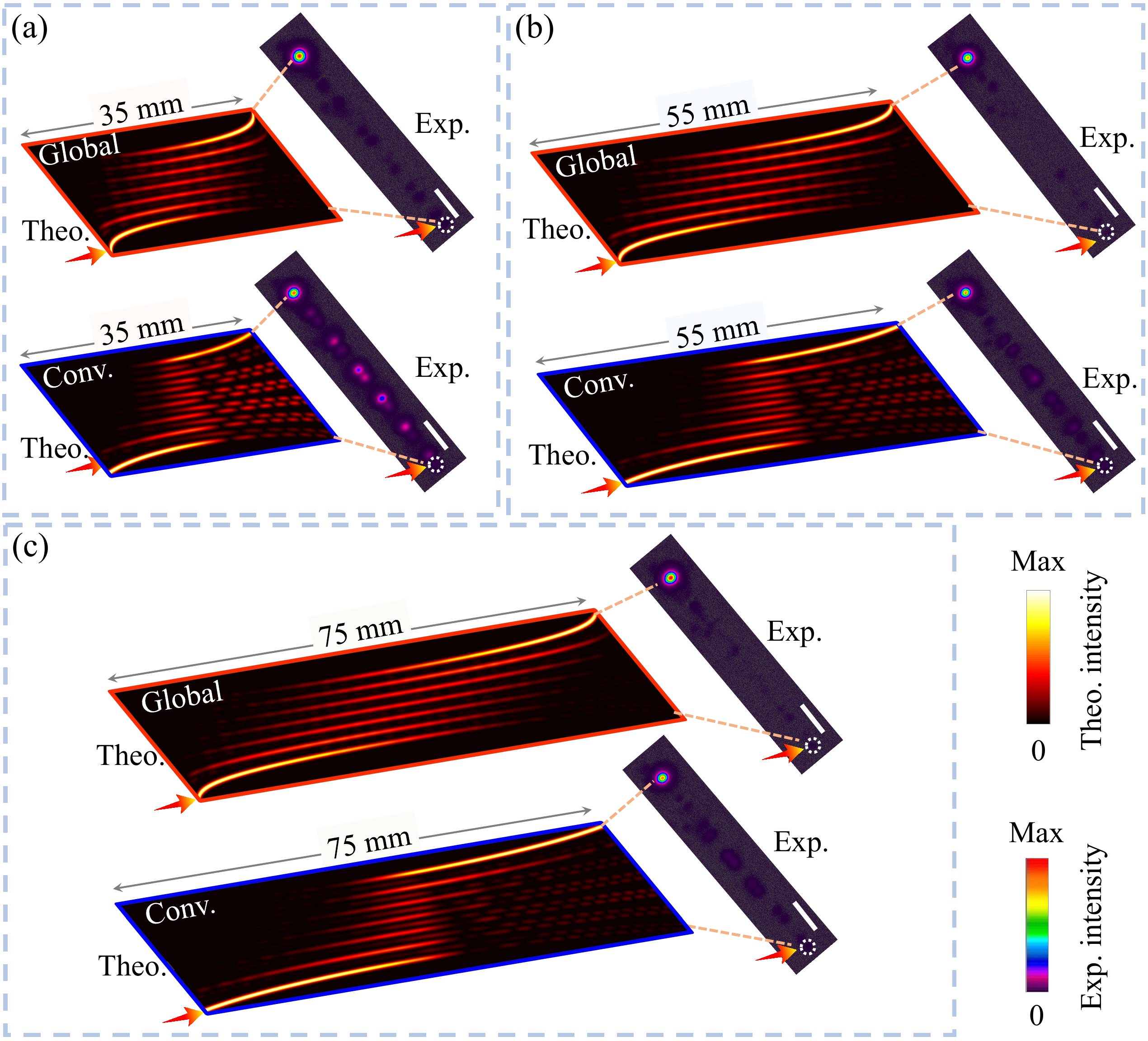}
		\caption{\textbf{Acceleration performance.}  (a)-(c)~Theoretical (Theo.) light propagation and experimentally measured (``Exp.") light intensity patterns at the output facet for our acceleration strategy and the conventional scheme with sample lengths $L=35$~mm, $55$~mm, and $75$~mm, respectively.}\label{f3}
	\end{figure}
	\textbf{\textit{Experimental demonstration}}---To validate the acceleration performance, we experimentally fabricated three sets of 13-waveguide samples with lengths $L=35$, $55$, and $75$~mm, respectively~(Note~5 in SM~\cite{SM}). Each set contains two samples corresponding to our acceleration strategy and the conventional scheme. An 808 nm wavelength laser source~(CNI, MDL-III-808L) was coupled into the first waveguide of each sample as the input port. The resulting light intensity patterns were captured at the output side of each sample using a charge-coupled device~(SP928, Ophir). Figures~\ref{f3}(a)-(c) (right panels) present the measured light intensity photographs of the end facets for the six samples. The measured output intensity distributions across the 13 waveguides in each sample are in excellent agreement with their respective theoretical values. According to both measured and predicted results, all the three samples for our acceleration strategy show almost fully localized light on the 13th waveguide representing the other edge of the waveguide array. On the contrary, for the conventional adiabatic pumping samples, the light intensity on the last waveguide is enhanced with increased sample length. Overall, significant light diffusion is observed in the bulk of the waveguide array of these conventional adiabatic samples, but our accelerated topological pumping sample of the same length exhibits strong light localization on the last waveguide. These results clearly demonstrate that our GAAC enables fast topological pumping for photonic transport with remarkably reduced device length and minimal bulk diffusion.
	
	\begin{figure}\centering
		\includegraphics[width=0.78\linewidth]{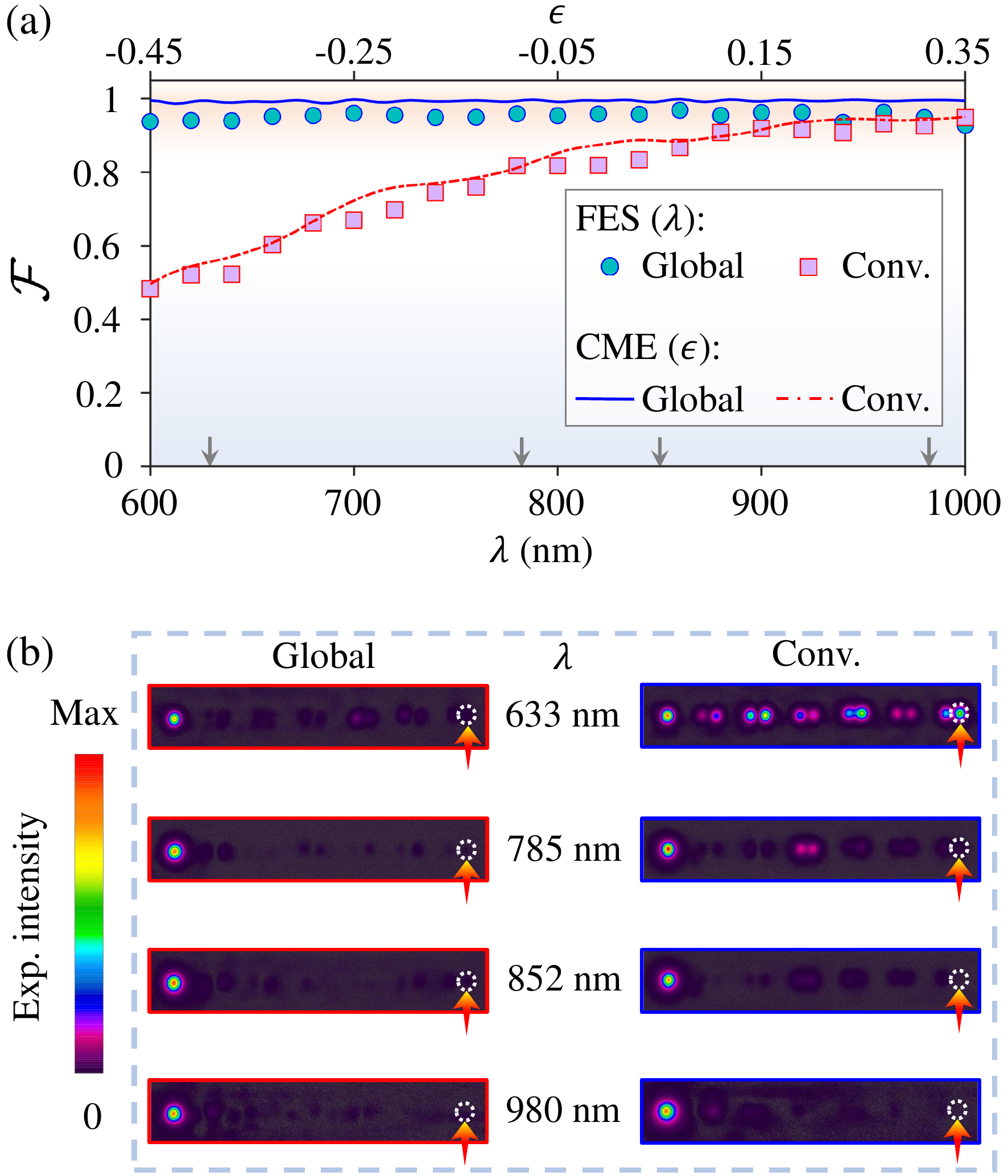}
		\caption{\textbf{Broadband performances.} (a)~Predicted pumping fidelity at $L=55$~mm with input wavelength ranging from $600$~nm to $1000$~nm for our acceleration strategy and the conventional pumping scheme. (b)~Measured light intensity patterns at the end facets of two samples working at specified four wavelengths instead of 808~nm.}\label{f4}
	\end{figure}
	To illustrate robustness of our acceleration strategy, we investigate its broadband performances by changing the input wavelength $\lambda\in[600,1000]$~nm to induce deviations in the inter-waveguide coupling strengths from their desired values. In Fig.~\ref{f4}(a), results predicted by finite-element simulations~(FES) show that our accelerated topological pumping maintains consistently a high fidelity across this entire spectral window, while the fidelity of the conventional adiabatic pumping increases with the input wavelength. Furthermore, to quantify effect of inter-waveguide coupling strength variations on the pumping fidelity, we further perform a CME analysis by introducing a relative deviation $\epsilon$ in Fig.~\ref{f4}(a), which perturbs the inter-waveguide coupling strengths $J_{1,2}(z)\rightarrow(1+\epsilon)J_{1,2}(z)$. This analysis exactly reveals a linear correspondence between the wavelength $\lambda\in[600,1000]$~nm and the deviation $\epsilon\in[-0.45,0.35]$~(Note~7 in SM~\cite{SM}). We experimentally corroborates these findings using four additional lasers with wavelengths (633, 785, 852, and 980 nm) coupled into the first waveguide of the two samples with $L=55$~mm. End-facet images in Fig.~\ref{f4}(b) show that our accelerated topological pumping maintains robust light confinement in the 13th waveguide across the tested spectra, while the conventional pumping scheme fails to sustain good localization at shorter wavelengths due to insufficient coupling strengths. These results confirm that our acceleration strategy exhibits robust broadband topological pumping performance even at a relatively short pumping length.

	\textbf{\textit{Conclusion}}---We have theoretically formulated and experimentally verified a high-efficiency framework for accelerating adiabatic topological pumping via GAAC, transcending the limitations of conventional instantaneous adiabatic conditions. By establishing that fidelity is fundamentally governed by the spatial mean ${\overline Q}_{\rm non}$ and variance $\sigma^2_Q$ of the nonadiabatic factor, we have facilitated rapid parameter modulation where the topological gap is large while preserving global adiabaticity. Using femtosecond-laser-written photonic lattices with scalable power-law couplings modulations, we have demonstrated high-fidelity, edge-to-edge transport with device footprint significantly shorter than those required by conventional pumping schemes. Our results exhibit the predicted linear scalability ($L \propto N$), and maintains exceptional robustness over a broad optical bandwidth ($>400$~nm). Moreover, as detailed by Note~6 in SM~\cite{SM}, our approach also surpasses other recently reported acceleration strategies~\cite{Song2024SciAdv,Wu2024NC,TTian2022PRL} in key metrics.
	
	The versatility of GAAC framework is further evidenced by its ready extension to functional devices, such as topological beam splitters (see SM, Note~8~\cite{SM}). While we have employed power-law coupling modulations as a primary realization, the GAAC is agnostic to the specific functional form, highlighting its broad applicability. This analytical GAAC can be synergistically integrated with numerical optimization methods~(e.g., the gradient-descent-based adiabaticity control limit~\cite{XLiu2025PRL}, time-optimal shooting method~\cite{Stepanenko2025}, or optimal-pulse genetic algorithm~\cite{WXLi2024PRA}) to systematically navigate the full design space. Such a synergy of analytical bounds and numerical refinement may provide a robust pathway for maximizing the speed and fidelity of topological transport, as well as broader adiabatic protocols, across diverse physical platforms from ultracold atoms~\cite{Leseleuc2019Science,TChen2024PRL,ZZhu2025PRL} to superconducting circuits~\cite{Deng2022Science,HQian2025Science,KZhou2025PRL}.

	\textbf{\textit{Acknowledgments}}---This work was supported in part by the National Natural Science Foundation of China under Grant Nos. 82003284 and 62375103, the National Key Research and Development Program of China under Grant Nos. 2024YFB4608100, 2021YFF0502700 and 2023YFB3208600. J.-L.W. acknowledges support from the National Natural Science Foundation of China (62571494,12304407). X.N. acknowledges support from the National Natural Science Foundation of China (12474396), and Key Project of Hunan Natural Science Foundation (2025JJ30001). Y.-K.S. acknowledges support from the National Natural Science Foundation of China (623B2042). K.-H.X. thanks X. Yang and W.-Z. Fan for their assistance with the sample fabrication and measurement.

	\textbf{\textit{Data availability}}---The data that support the findings of this article are openly available~\cite{khXiao2025}.

	\bibliography{myref}~
	
\end{document}